\documentclass[twocolumn,pre,showpacs]{revtex4}
\usepackage{graphicx}
\usepackage{psfrag}
\usepackage{array}
\usepackage[intlimits]{amsmath}
\usepackage{amssymb}
\usepackage{xspace}

\newcommand{\varh}[1]{\ensuremath{\langle{}#1{}\rangle}}
\newcommand{\atlag}[1]{\overline{#1}}
\newcommand{\llim}[1]{\ensuremath{\underset{#1}\lim}\xspace}

\newcommand{\rmax}{\ensuremath{r_{\text{max}}}\xspace}
\newcommand{\rn}{\ensuremath{r_{\text{n}}}\xspace}
\newcommand{\intreg}{\ensuremath{\mathfrak{A}}\xspace}
\newcommand{\limN}{\llim{N\to\infty}}
\newcommand{\figref}[1]{Fig.~\ref{#1}\xspace}
\newcommand{\secref}[1]{Section~\ref{#1}\xspace}
\newcommand{\CNL}{Communication Networks Laboratory, E\"otv\"os University, Budapest, Hungary}

\begin{document}
\title{Giant Clusters in Random Ad Hoc Networks} 
\author{G.~N\'emeth}
\email{gab@cnl.elte.hu}
\affiliation{\CNL}
\author{G.~Vattay}
\affiliation{\CNL}
\date{\today}
\begin{abstract}
The present paper introduces ad hoc communication networks as examples of
large scale real networks that can be prospected by statistical means. A
description of giant cluster formation  based on the single
parameter of node neighbor numbers is given along with the discussion of
some asymptotic aspects of the giant cluster sizes.
\end{abstract}
\maketitle

\section{Introduction}\label{sec:intro}

Nowadays, natural and designed networks are in the focus of research in different
scientific 
disciplines. Using computers the amount of available empirical data on real 
world networks has been increased during the past few years. 
Examples of real networks include the 
World Wide Web\cite{DiameterOfTheWWW,Broder},
the Internet 
\cite{falou,JB,Cohen_error,Cohen_attack,Shavitt}, 
collaboration networks of movie actors
and  scientists
\cite{Barab_collab,Newman_collab1,Newman_collab2},
power grids
\cite{WattsStrogatz,Watts_book}
 and the metabolic network of  living organisms
\cite{Metabolic,WagnerFell,SoleMontoya1,SoleMontoya2}. 

Random graphs are natural candidates for the description
of the topology of such large systems of
similar units. 
In \cite{ER1,ER2,ER3} the authors have developed a model
-- which assumes each pair of the graph's vertices
to be connected with equal and independent probabilities --
that treats a network as an assembly of {\it equivalent units}. 

This model, introduced by the mathematicians 
Erd\H os and R\'enyi,
has been much investigated in the mathematical literature\cite{Bollobas,BollobasThomason85}.
However, the increasing availability of large maps of real-life 
networks has indicated that the latter structures are fundamentally correlated
systems, and in many respects their topologies deviate from the
uncorrelated random graph model.

Two classes of models, commonly called the {\it small-world
graphs}\cite{WattsStrogatz,Watts_book,Amaral_crossover}
and the {\it scale-free networks}\cite{EmergenceOfScaling,ScaleFree_PA} have been
developed to capture the clustering and the power law degree distribution present 
in real
networks\cite{Redner,WattsStrogatz,Watts_book,falou,AdamicHuberman,DiameterOfTheWWW,Metabolic,clever,%
EmergenceOfScaling,ScaleFree_PA,Amaral_ClassesOfSW,Barab_collab,Newman_collab1,Newman_collab2,fij}.

\begin{figure}
\begin{center}
\psfrag{s}[c][c]{s}
\psfrag{d}[c][c]{d}
\includegraphics[keepaspectratio,width=.5\linewidth]{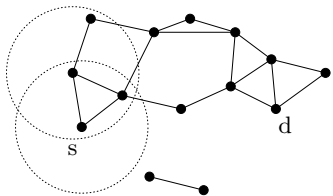}
\end{center}
\caption{Nodes and connections of an example ad hoc network. The
transmission range is the same for all nodes -- it is noted by the dotted
circle for two of the nodes. The shortest
path between the s source and d destination users touches 3 intermediate nodes,
and there is an alternative route of 6 hops, which has no common intermediate nodes with the
first.}
\label{fig:anet}
\end{figure}

Here we  present \emph{ad hoc networks}\cite{manet} as new examples of real structures that can 
be investigated using the above network models. Ad hoc networks arise in
next generations of communication systems and hereby we try to summarize the
principal characteristics of such systems. In the ad hoc scheme users communicate
by means of short range radio devices, which means that every device can
connect to those devices that are positioned no farther than a 
finite maximum geometrical range. We call this range the given device's \emph{transmission
range} and the exact value of this range may depend on the transmitter's power and various
other physical parameters. See \figref{fig:anet} for an example ad hoc
network topology. Neighbor nodes talk the way ordinary radios -- like CBs -- do, however communication between non-neighboring
users is also possible. The latter case
is accomplished by sending the information from the source user to the destination hop by
hop, through intermediate nodes.
If the density of users in the area is high compared to their transmission
ranges, it is highly possible that more than one alternative route exists between
two users. This last feature can be exploited in the case if the shortest
route is overloaded or broken, or if the system allows  splitting the
information flow into separate parallel flows.
Moreover, the users are free to move randomly and organize themselves arbitrarily; thus, the network's
 topology may change rapidly and unpredictably. Such a network may operate in a stand-alone fashion, or
 may be connected to the Internet. 

Giant clusters in ad hoc networks are made interesting because a communication network provides a
meaningful service only if it integrates as many users as possible within the covered area
(e.g., 99\% may be considered a good coverage).
In this paper we introduce a fractal model, that duplicates the giant component formation in ad hoc networks in an area inlaid
with obstacles, partially screening radio transmission. 
Our main result is that in such  networks the giant component
size can be described by a single parameter: the average number of
neighbors a node has.
The rest of this paper is structured as follows.  \secref{sec:model}
gives a detailed description of our random ad hoc network model.
In Sections~\ref{sec:rand} and \ref{sec:fractal} we delve into the topology differences between
random graphs and graphs built using our model.
\secref{sec:sim}  shows the numerical simulation results  supporting these
analyses. 

\section{The Random Ad Hoc Network Model}\label{sec:model}

A wireless ad hoc network consists of a number of
radio devices, also referred to as "nodes" in the following. Every node may
be connected to one or more other nodes in her vicinity; the actual set of
connections depends on the distance of the nodes. In a static environment
these connections define the topology of the system; if the nodes allowed to
move then the topology may change, however at any given point of time there
is still a well defined topology available.

To be precise we define a \emph{random ad hoc network} as a set of 
uniformly distributed nodes on the arena of the unit Euclidean square \([0;1]\times[0;1]\) with connections
between pairs of them. The connections are two-way in the sense that if node
\(A\) can communicate to node \(B\), then node \(B\) is also able to
communicate to node \(A\). 

Two nodes are connected if the geometrical distance of the two is less than
a certain value \(r_t\), that is the nodes can communicate up to their "transmission
range". 
We represent a realization of such a system using an undirected graph \(G(V,E)\), 
where the vertices and the edges denote the nodes and the two-way connections
respectively.
Sometimes a graph resulting this way is referred to as a
\emph{geometric random graph} or \emph{GRG}.
Note that there are no loops and no multiple edges in \(G\): \emph{a)} a node should not
communicate to itself; and \emph{b)} if two nodes are neighbors, then
technically there is no sense  to open a second communication channel
between them.

Furthermore, all the length parameters in the system are made dimensionless as follows.
Length is measured as  the multiples of the
\emph{unit radius} \(r_0\), which is in turn defined by the share of the whole area for
each node:
\begin{equation}\label{eq:r0}
r_0:=\sqrt{\frac A{N\pi}}
\end{equation}
where \(A\) is the size of the arena.
The ratio of the transmission range and the unit radius is called
the \emph{normalized transmission range} and noted by:
\begin{equation}
\rn:=\frac {r_t}{r_0}
\end{equation}

As mentioned in the Introduction, a communication network may deliver
meaningful service only if the network is connected, or at
least has a vast subset that is connected. Our work is focused on
examining the criteria for giant cluster formation and in particular in
networks with fractal connectivity properties. 

In the following we give a short overview of networks
on random graphs and afterwards we turn to our model of fractal ad hoc connectivity.

\section{Connectivity in Random Networks}\label{sec:rand}

After distributing and connecting the nodes as described previously,
the largest connected component of \(G\) can be determined. Let \(S\) be this components' size fraction:
\[S:=\frac{\text{nodes in the largest component}}N\] which quantity is obtained by counting.
This quantity is of particular importance because the network gets fully  connected if \(S\) diverges and for this end
we are to investigate its relationship with other network parameters.

In \cite{strogatz} the authors present the theory of \emph{random graphs}\/\cite{ER2} of arbitrary degree distribution. Among others, an exact result for 
the component sizes is given, which we shall cite here.
It is shown, that the average component size diverges if \[\sum_{k=0}^{\infty}k(k-2)p_k=0\] holds, where \(p_k\) is the degree
distribution of vertices in \(G\). Let us use here the actual distribution of our ad hoc network:
it is easily seen that the probability distribution of the number of nodes
contained in any disc with radius \rn
is the Poisson-distribution with expectation value of \(\rn^2\). It means that 
\begin{equation}
p_k=\frac{\left(\rn^2\right)^k}{k!}e^{-\rn^2}
\label{eq:pk}
\end{equation} is
the probability that a vertex will have \(k-1\) neighbors (the \(-1\) is
because the node itself does not count for a neighbor). Applying the result in \cite{strogatz}, one can derive
the relationship of the size of the giant component \(S\) and the
transmission range:
\begin{equation}
\rn^2=\frac{\log(1-S)}{-S}
\label{eq:rs}
\end{equation}
It shall be noted here that while (\ref{eq:rs}) holds for random networks and --
as it is to be shown in \secref{sec:sim} -- for fractal ad hoc networks, the
\(\rn-S\) relationship is different for the finite range ad hoc case,
however, the latter is to be discussed in a separate paper.

\section{The Fractal Ad Hoc Neighborship Algorithm}\label{sec:fractal}

The results of the previous section apply for scenarios where the arena is
"flat": that is the only limit to build a connection between two nodes is
their geometrical distance.
In the present section we introduce the idea of generalized obstacles that
can screen nodes from each other even if they are positioned within transmission range.
This change produces graphs with extended spatial structure which is why we
call the algorithm fractal.

The obstacles are adopted by changing 
the algorithm for edge generation. Now two nodes within the transmission range will be connected with a
probability which is given as the function of their geometrical distance.
For every two nodes
\(u,v\in V\) let
\(p({\mathrm dist}(u,v))\) be the probability that an edge \(e_{uv}\in E\)
connecting them is set up.
For the description of our obstacles we use a long tailed probability
function
which is implied by  the picture of a hilly landscape, where the
possibility of connections drops with the increasing geometrical distance
between the nodes, however long range connections are  still possible:
\begin{equation}\label{eq:pr}
p(r)=\frac a{\left(1+\frac r{r_0\beta}\right)^{\beta}}
\end{equation}
with parameter values \(a>0\) and \(\beta>0\).

\begin{figure}
\begin{center}
\psfrag{N}[c][c]{\(N\)}
\psfrag{S}[r][r]{\(S\)}
\psfrag{b3.5}[c][c]{\(\beta=3.5\)}
\psfrag{b3.0}[c][c]{\(\beta=3.0\)}
\psfrag{b2.5}[c][c]{\(\beta=2.5\)}
\psfrag{b2.0}[c][c]{\(\beta=2.0\)}
\psfrag{b1.5}[c][c]{\(\beta=1.5\)}
\psfrag{b1.0}[c][c]{\(\beta=1.0\)}
\includegraphics[keepaspectratio,width=\linewidth]{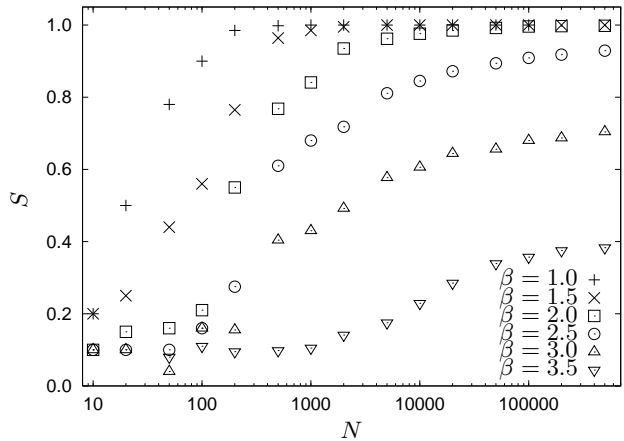}
\end{center}
\caption{Giant component sizes for various values of \(N\) and \(\beta\). Note that \(S(N)\) reaches \(1\) for
\(\beta\leqslant2\), yet \(\limN S(N,\beta)=S_{\text{max}}(\beta)<1\) for \(\beta>2\).}
\label{fig:sn}
\end{figure}

Performing computer simulations of networks connected according to (\ref{eq:pr}), one obtains different results, as \(\beta\) changes. On \figref{fig:sn}
we compared the resulting giant cluster sizes for different \(\beta\) values. At
lower parameter values \(S(N)\) saturates to \(S=1\) -- all nodes become elements of the giant cluster above a certain finite node number.
For \(\beta=2.5\) and above \(S\) still converges to a finite value, however the limit now is strictly less than \(1\). It means that
networks with such parameter values will not become fully connected even at large node numbers, moreover, the proportion of the largest connected
subgraphs drops with \(\beta\) worse than linearly.
In the rest of this Section we try to interpret this dual behavior of \(S(\beta)\).

It is easy to imagine that the more
connections the nodes have in average, the larger the giant cluster grows. More accurately we state that the
\emph{average vertex degree} \(\varh C\) determines the cardinality of the largest connected subgraph in \(G\). Clearly, if \(\varh
C=0\), then every connected component contains a single node, and in the \(N\to\infty\) limit \(S\) becomes \(0\). Also, if
\varh C diverges or even if only a single node is connected to all the others, the graph obviously gets fully connected. Based on these
considerations we are to examine \(\varh C\) in detail.

Vertex degree in \(G\) can be calculated by fixing a single
node and totaling 
the \(\varh{C_r}\) expectation value of  the number of neighbors that reside
exactly at the distance \(r\) away from the fixed one.
 Assuming that the density of nodes is constant (\(N/A\)),
\(\varh{C_r}\) can be expressed by multiplying the average number of nodes
in distance \(r\) and the probability (\ref{eq:pr}):
\[
\varh{C_r}=\frac{2r\pi}{A}\,N\,p(r)
\]
Now if \(\atlag\rho=N/A\), the average vertex degree is
\begin{equation}\label{eq:Cint}
\varh C=\int_\intreg \varh{C_r}\,dr
=\int_\intreg p(r)2\pi\atlag\rho r\,dr
\end{equation}
where \intreg  represents the physical boundaries of
the arena. As there are no  nodes outside this region, thus the integral
shall be \(0\) outside \intreg.

In general solving (\ref{eq:Cint}) yields
\begin{multline}\label{eq:Cgen}
\varh C
=a2\pi\atlag\rho\frac{r_0\beta}{1-\beta}\\
\times\left[r\left(1+\frac 
r{r_0\beta}\right)^{1-\beta}-\frac{r_0\beta}{2-\beta}\left(1+\frac
r{r_0\beta}\right)^{2-\beta}\right]_\intreg
\end{multline}
However the expectation value of \(\varh C\)
is dependent on the value of \(\beta\).
Accordingly, our discussion is separated into several cases.

\begin{itemize}
\item[a]\(\beta>2\).
In this case  (\ref{eq:Cgen}) can be evaluated for
\intreg being the interval \(r\in[0;\infty)\) in the limit where \(r_0\to0\):
\begin{equation}\label{eq:>2}
\varh C=
\frac{a2\pi\atlag\rho\cdot r_0^2\beta^2}{(1-\beta)(2-\beta)}
\equiv\frac{2a\beta^2}{(\beta-1)(\beta-2)}
\end{equation}
Furthermore, knowing that 
\[
\lim_{\alpha\to\infty}\frac1{(1+x)^\alpha}=e^{-\alpha x}
\]
in the \(\beta\to\infty\) limit (\ref{eq:>2}) becomes 
\[
\varh C=a2\pi\atlag\rho r_0^2\equiv2a
\]

\item[b]\(\beta=1\) or \(\beta=2\).
(\ref{eq:Cgen}) diverges logarithmically in \(r\), thus \(C\) does
not have an expectation value.

\item[c]\(\beta<2\) and \(\beta\neq1\).
Here \(\varh C\) will diverge as \(N\to\infty\), however
unlikely to the previous case
we  try to determine the \(\varh{C(N)}\) relation.
First let us rewrite (\ref{eq:Cgen}) as
\begin{equation}\label{eq:rewr}
\varh C=\frac{a2\pi\atlag\rho\cdot r_0\beta}{(1-\beta)(2-\beta)}
\left[\frac{r(1-\beta)-r_0\beta}
{(1+\frac r{r_0\beta})^{\beta-1}}
\right]_\intreg
\end{equation}

Concerning the \(r\)-dependence in \([\ldots]_\intreg\) we can assume that 
there is a maximal transmission range \rmax such that for transmission  ranges
\(r>\rmax\) the contribution of the integrand in (\ref{eq:Cint}) is
negligible.
This way the \([\ldots]_\intreg\) part of (\ref{eq:rewr}) can be estimated as 
\begin{equation}\label{eq:rmax}
\left[\ldots\right]_\intreg\simeq-\frac{r_0\beta}{\left(1+\frac r{r_0\beta}\right)^{\beta-1}}
+\frac{\rmax\cdot(1-\beta)}{\left(1+\frac {\rmax}{r_0\beta}\right)^{\beta-1}}
\end{equation}
Now if \(r_0\to0\) (which happens to be the case at  sufficiently large  node numbers) the first term
in (\ref{eq:rmax}) vanishes and the \(+1\) becomes negligible in the 
denominator of the second term.
After substituting this second term 
 and  simplifying the expression, (\ref{eq:Cgen}) finally becomes
\[
\varh C\simeq\frac{a2\pi\atlag\rho}{2-\beta}
\left(\frac{r_0\beta}{\rmax}\right)^{\beta}\cdot\rmax^2
\]

The \(N\)-dependence of \(\varh C\) can be derived from here by substituting
definition (\ref{eq:r0}), \(\atlag\rho=N/A\) and the fact that
\(\rmax^2\propto A\). By these means the above expression yields:
\begin{equation}\label{eq:Cest}
\varh C\propto N^{1-\frac{\beta}2}
\end{equation}
\end{itemize}

To summarize, if \(\beta>2\), then a finite neighbor count is expected, and thus such networks are not going to be fully connected
(see again \figref{fig:sn}). On the other hand, if \(\beta<2\), then 
 \varh C diverges exponentially with increasing node numbers, which in theory leads to fully connected networks at large \(N\),
and means, that the more nodes are in the system, the larger the fraction of connected nodes is to become.

\section{Simulation Results}\label{sec:sim}
We carried out computer simulations to illustrate  our  findings, especially Eqs.\ (\ref{eq:>2}) and (\ref{eq:Cest}). 
During a simulation run we first  pick the random coordinates for the \(N\)
 nodes.  Second the probability
\(p\) is calculated  according to (\ref{eq:pr}),
using the input parameters \(a\),\(\beta\) and \(r\).
 Then for every two nodes a uniform random number \(\xi\in[0;1]\)
is generated and compared to \(p\): for cases \(\xi<p\) an edge 
connecting those two nodes is recorded. Finally we count the component sizes and take the largest of these. 
The output of the simulation run is the average vertex degree, \(\varh C\), and the largest components' size, \(S\).
Note the analogy with (\ref{eq:pk}) if using the fact, that here \(\varh C=\rn^2\).

\begin{figure}
\begin{center}
\psfrag{S}[c][c]{\(S\)}
\psfrag{C}[c][c]{\(\varh C\)}
\psfrag{b=2.5}[r][r]{\(\beta=5.6\)}
\psfrag{b=1.56}[r][r]{\(\beta=1.56\)}
\psfrag{random graphs}[r][r]{random graphs, Eq.~(\ref{eq:rs})}
\includegraphics[keepaspectratio,width=\linewidth]{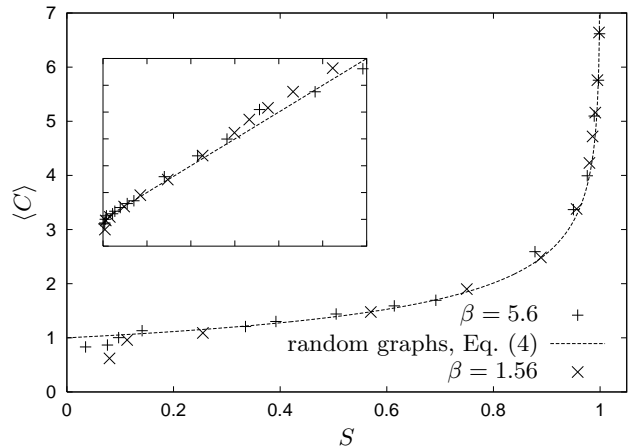}
\end{center}
\caption{Simulations of ad hoc networks using the fractal neighborship algorithm with
parameter values both \(\beta<2\) and \(\beta>2\) yield the same giant
cluster size vs.\ average vertex degree as random graphs. Inset displays
the same plots with \(-\log(1-S)/S\) as the abscissa.}
\label{fig:cs}
\end{figure}

\begin{figure}
\begin{center}
\psfrag{N}[c][c]{\(N\)}
\psfrag{C}[c][c]{measured \(\varh C\)}
\includegraphics[keepaspectratio,width=\linewidth]{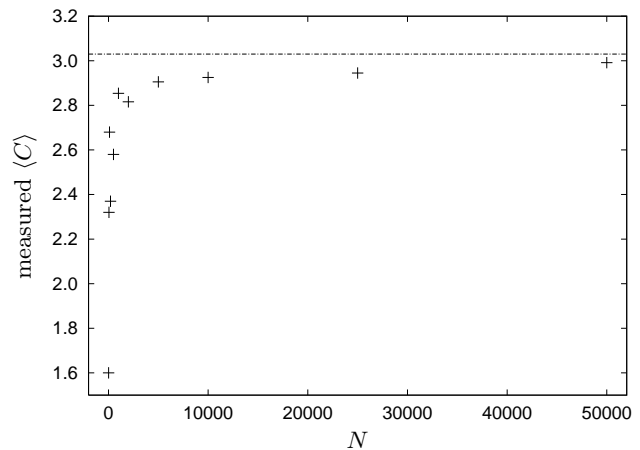}
\end{center}
\caption{Average vertex degree of ad hoc graphs for \(\beta>2\). Data points
were acquired using \(a=0.8\), \(\beta=5.6\);
the dashed line yields the analytical result \(\varh C=3.03\), which 
shall hold  in the \(N\to\infty\) limit.}
\label{fig:cn_5.6}
\end{figure}

As the first test we recorded the giant cluster size vs.\ transmission range
relationship.
Data points were obtained by repeated runs,  changing only the amplitude
parameter \(a\) of (\ref{eq:pr}) in an appropriate interval (e.g.,
\(a\in[1;9]\) for the \(\beta=2.5\) case).
\figref{fig:cs} illustrates that
in a network connected using
the fractal neighborship algorithm the observable \(S-\varh C\) relationship
matches the equivalent analytical result for random
graphs for both relevant cases \emph a and \emph c in \secref{sec:fractal}.

On the other hand, the behavior of \(\varh C\) turns out to be sensible to the value of
\(\beta\), as it was expected. Let us  start with the case  \(\beta>2\).
\figref{fig:cn_5.6} presents the simulation
results for networks connected as by (\ref{eq:pr}), using \(a=0.8\) and
\(\beta=5.6\). According to (\ref{eq:>2}), the average vertex degree
is expected to be
\[
\varh C=\frac{2\cdot 5.6^2}{4.6\cdot3.6}\simeq3.03
\]
in this case. It is clearly seen on the Figure that increasing \(N\),
the simulation output converges to the analytical result.

\begin{figure}
\begin{center}
\psfrag{N}[c][c]{\(N\)}
\psfrag{C}[c][c]{measured \(\varh C\)}
\includegraphics[keepaspectratio,width=\linewidth]{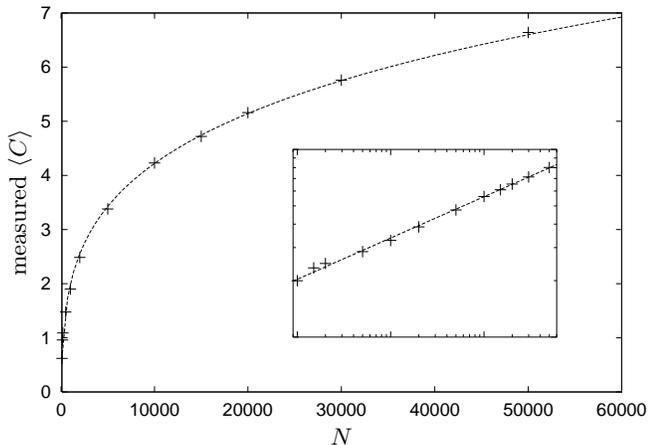}
\end{center}
\caption{Divergence of average vertex degrees with \(N\) for \(\beta<2\).
Crosses: data points for \(a=0.1\), \(\beta=1.56\); dashed line is
\(\propto N^{1-\beta/2}=N^{0.22}\). Inset displays the same plot with both axes
logarithmic.}
\label{fig:cn_1.56}
\end{figure}

Now let us turn to the \(0<\beta<2\) case. On \figref{fig:cn_1.56} the
data obtained for \(a=0.1\) and \(\beta=1.56\) is shown along with a numeric
function fit according to (\ref{eq:Cest}):
\[
\varh{C(N)}=c_0\cdot N^{1-\frac{1.56}2}+c_1
\]
(the parameters turn out to be \(c_0=0.74\) and \(c_1=-1.38\)). The
simulations agree with the \(N^{1-\beta/2}\) divergence well, as calculated in 
\secref{sec:fractal}. 

Figs.~\ref{fig:cn_5.6} and \ref{fig:cn_1.56} now illustrate the differing \(S\)-behavior presented on \figref{fig:sn}, as a data set
for \(\beta=5.6\) would converge to some connectivity \(<20\%\) even at very large \(N\), while the one for \(\beta=1.56\) is clearly reaching \(S=1\) for node
numbers in the magnitude of several thousands.

\section{Conclusions}\label{sec:concl}
In the present paper we have investigated the  connected components that are produced in random ad hoc networks.
Based on the results, the number of nodes needed for a given connectivity ratio can be estimated.
Thus our results may hint about the usefulness of random fractal ad hoc networks.

We modified the conventional connection function and made long range connections possible. This way the producing
networks become extended in their spatial structure, as thought the network is situated in an area with obstacles
screening some of the transmissions.
We have found that a single parameter -- the average neighbor count \(\varh C\) -- can characterize the proportion of the largest connected
subnetwork. We have also seen that depending on the connection function parameters, this proportion can be either bounded or
unbounded as the system size \(N\) is increased. For both cases \(\varh{C(N)}\) was derived analytically and confirmed by simulations.

\section*{Acknowledgements}
The authors thank the support of Hungarian National Science Found
OTKA T37903 and T32437.

\bibliography{cprops}

\end{document}